\documentclass[bibtex]{llncs}

\usepackage{color}
\usepackage{amssymb}
\usepackage{amsmath}
\usepackage{graphicx}
\usepackage{subfigure}
\usepackage{units}
\title{Complex synchronization patterns in the human connectome
  network}  \titlerunning{Complex synchronization patterns}

\author{Pablo Villegas$^1$, Jorge Hidalgo$^1$, Paolo Moretti$^2$, and Miguel A. Mu\~noz$^1$} 

\authorrunning{P. Villegas et al.}

\institute{$^1$ Departamento de Electromagnetismo y F{\'i}sica de la
  Materia e Instituto Carlos I de F{\'i}sica Te{\'o}rica y
  Computacional.  Universidad de Granada, E-18071 Granada,
  Spain. \newline $^2$ Institute of Materials Simulation (WW8),
  Friedrich-Alexander-University Erlangen-N\"unberg,
  Dr.-Mack-Stra{\ss}e 77, D-90762 F\"urth, Germany.}

\begin{document}

\maketitle

\begin{abstract}
  A major challenge in neuroscience is posed by the need for relating
  the emerging dynamical features of brain activity with the
  underlying modular structure of neural connections, hierarchically
  organized throughout several scales. The spontaneous emergence of
  coherence and synchronization across such scales is crucial to
  neural function, while its anomalies often relate to pathological
  conditions.  Here we provide a numerical study of synchronization
  dynamics in the human connectome network. Our purpose is to provide
  a detailed characterization of the recently uncovered broad dynamic
  regime, interposed between order and disorder, which stems from the
  hierarchical modular organization of the human connectome.  In this
  regime --similar in essence to a Griffiths phase-- synchronization
  dynamics are trapped within metastable attractors of local
  coherence.  Here we explore the role of noise, as an effective
  description of external perturbations, and discuss how its presence
  accounts for the ability of the system to escape intermittently from such
  attractors and explore complex dynamic repertoires of locally
  coherent states, in analogy with experimentally recorded patterns of
  cerebral activity.
  
\end{abstract}


\section{Introduction}\label{s1}

The current mapping of neural connectivity patterns relies on advanced
neuro-imaging techniques, which have recently allowed for the
reconstruction of structural human brain networks, establishing at an
individual-based level which brain regions are mutually connected, as
well as the strength of pairwise connections.  The resulting ``human
connectome network'' \cite{Hagmann,Honey09} has been found to be
structured in moduli or compartments --characterized by a much larger
intra than inter connectivity-- organized in a hierarchical
fractal-like fashion across diverse scales
\cite{Bullmore-Sporns,Sporns,Review-Kaiser,Review-Bullmore,Zhou06,Ivkovic}.
On the other hand, functional connections between different brain
regions can be inferred e.g. from correlations in neural activity as
detected in EEG or fMRI time series.  Unveiling the relation between
structural and functional networks is a current challenge in modern
neuroscience. In this context, a few pioneering works found that the
hierarchical-modular organization of structural brain networks has
remarkable implications for neural dynamics
\cite{Zhou06,Zhou07,Kaiser07,Kaiser10}. As opposed to the case of
simpler network structures, neural activity propagates in hierarchical
networks in a peculiar way.  For example, models of neural activity
propagation usually exhibit two familiar phases --percolating and
non-percolating, respectively-- but it has been recently found
\cite{Nat-Comm} that when such models operate on top of the ``human
connectome'' structural network a novel intermediate regime, named a
``Griffiths phase'' \cite{Vojta-Review,GPCN} emerges.  This novel
phase originates from the highly-diverse and relatively isolated
structural moduli where dynamical activity may remain mostly localized
for long time periods \cite{Nat-Comm,GPCN}.

Given that the correct brain functioning requires coherent neural
activity at a wide range of scales \cite{Niebur2000,Kandel00}, the
study of synchronization among neural populations is one of the
central ideas in computational neuroscience
\cite{Buzsaki,Breakspear-multiscale}.

In a recent work \cite{Villegas2014}, some of us scrutinized the
special features of synchronization dynamics \cite{RPK-book} using the
canonical Kuramoto model for phase synchronization
\cite{Kuramoto75,Strogatz00,Acebron-Review}, in the actual human
connectome (HC) network \cite{Hagmann,Honey09,Arenas-Review}. In
analogy to what described above for activity propagation, we uncovered
the existence of a novel intermediate phase for synchronization
dynamics, stemming from the hierarchical modular organization of the
HC. Furthermore, we found that the dynamics in such a region presented
a plethora of complex and interesting dynamical features
\cite{Villegas2014}.

Our goal here is to describe in more detail the complex behavior
within such an intermediate regime, both in individual moduli and at a
global brain level.  We measure the fluctuations of the global order
parameter as a function of the overall coupling strength, and we show
that there is a broad region (rather than a unique ``critical'' point)
with huge variability and response. Finally, we assess the role of
noise and perturbations in the robustness of the metastable stated
arising in the intermediate regime, and we show that adding intrinsic fluctuations to the picture of
synchronization dynamics in hierarchical modular networks accounts for
the ability of the brain to explore different attractors, giving
access to the varied functional configurations recorded in experiments
\cite{Chialvo10,Deco2012,Haimovici}.

\section{Kuramoto model in the Human-Connectome network}

The HC network we employ consists of a set of $N=998$ nodes, each of
them representing a population of neurons producing self-sustained
oscillations \cite{Sporns2011}, connected pairwise through a precise
pattern of symmetric weighted edges, altogether determining a
connectivity matrix $\mathbf{W}$ \cite{Hagmann,Honey09}.

On top of such a HC network, we implement a noisy Kuramoto dynamics,
defined by the set of differential equations \cite{Kuramoto75,Strogatz00,Acebron-Review}:
\begin{equation}
{\dot\theta}_{i}(t)=\alpha \eta_i(t)+\omega_{i}+ k \sum_{j=1}^{N}
W_{ij}\sin\left[\theta_{j}(t)-\theta_{i}(t)\right], \label{eq:Kuramoto}
\end{equation}
where ${\theta}_{i}(t)$ is the phase at node $i$ at time $t$, the
intrinsic frequencies $\omega_{i}$ --accounting for region
heterogeneity-- are extracted from some probability distribution function $g(w)$,
$W_{ij}$ are the elements of the $N \times N$ weighted connectivity matrix
$\mathbf{W}$, $k$ is an overall coupling parameter and $\eta_i(t)$ is a
zero-mean delta-correlated Gaussian noise, tuned by the real-valued
amplitude $\alpha$.

The Kuramoto complex order parameter is defined by
$Z(t)=R(t)e^{i\psi(t)} = \langle e^{i \theta_k(t)} \rangle_k$, where $0
\leq R \leq 1$ gauges the overall coherence and $\psi(t)$ is the
average phase.  It is common wisdom that for an (infinitely) large
population of oscillators interacting in a fully connected network,
the model exhibits a phase transition at some value of $k$, separating
a coherent steady state ($R > 0$) from an incoherent one ($R=0$, plus
$1/\sqrt{N}$ finite-size corrections)
\cite{Kuramoto75,Strogatz00,Acebron-Review}. On the other hand, in the
absence of frequency heterogeneity the system always reaches a
coherent state. Thus, frequency heterogeneity is responsible for
frustrating synchronization if the coupling strength is weak.
Similarly, in our recent work \cite{Villegas2014} we argued that the
combined effect of frequency heterogeneity {\it and} network
heterogeneity (in particular, a hierarchical modular structure) can
lead to much richer and interesting ways of ordering frustration.
Here we explore that phenomenology in much deeper detail, introducing
external stochastic fluctuations (i.e. noise) as the mechanism
accounting for the ability of the system to explore metastable
configurations.

\begin{figure}[h]
 \centering\includegraphics[width=0.7\textwidth]{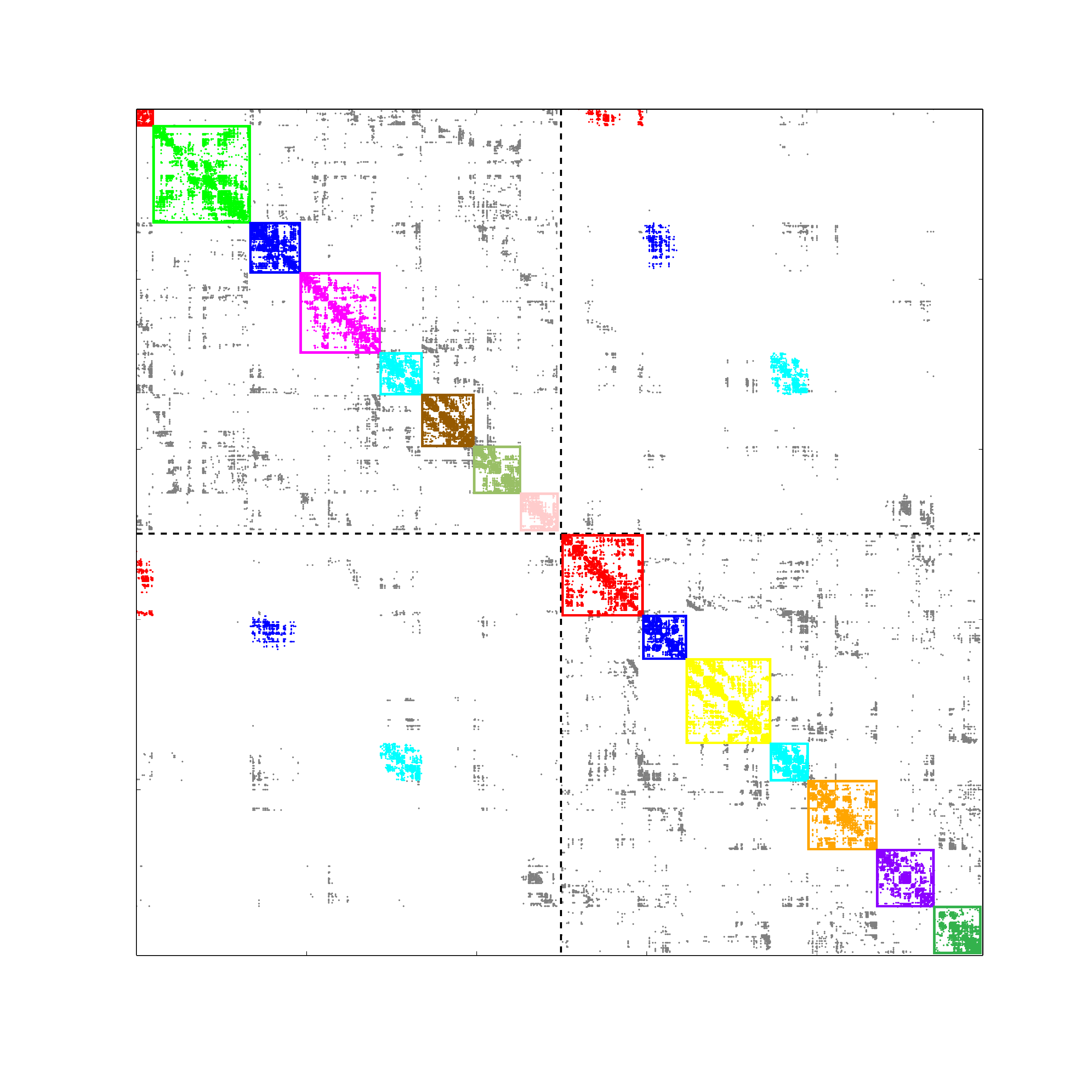}
 \caption{Adjacency matrix of the HC network with nodes ordered to
   emphasize its modular structure as highlighted by a community
   detection algorithm (see main text), showing also the partition
   into the $2$ hemispheres (dashed lines). $12$ moduli can be
   distinguished (each plotted with a different color); $4$ of them
   correspond to one of the two hemispheres, $5$ to the other, and
   only $3$ moduli overlap with both hemispheres (cyan, blue and red
   moduli).  Inter-modular connections (grey) are limited to small
   subsets, acting as interfaces or connectors between moduli.}
\end{figure}

\section{Results}

We considered the HC network \cite{Hagmann,Honey09} and employed
standard community detection algorithms \cite{Radatools,Ivkovic} to
identify the underlying modular structure.  The optimal partition into
communities --i.e. the one maximizing the modularity parameter
\cite{Newman-Review}-- turns out to correspond to a division in $12$
moduli \cite{Villegas2014}. At a higher hierarchical level, a
separation into just $2$ moduli (roughly corresponding to the $2$
cerebral hemispheres) also provides a quite high modularity value.  As
illustrated in Fig. 1, $4$ (out of the $12$) moduli belong to one of
the two hemispheres, $5$ to the other, while $3$ moduli (cyan, blue
and red) overlap with both hemispheres.  We label these two
hierarchical levels as $l=2$ ($2$ large moduli) and $l=1$ ($12$
smaller moduli), respectively.

We have conducted computational analyses of the noisy Kuramoto model
on top of the HC network and performed a number of new computational
experiments complementing the analyses in our previous work
\cite{Villegas2014}.

\begin{figure}[h]
\centering
\includegraphics[width=9cm,angle=0]{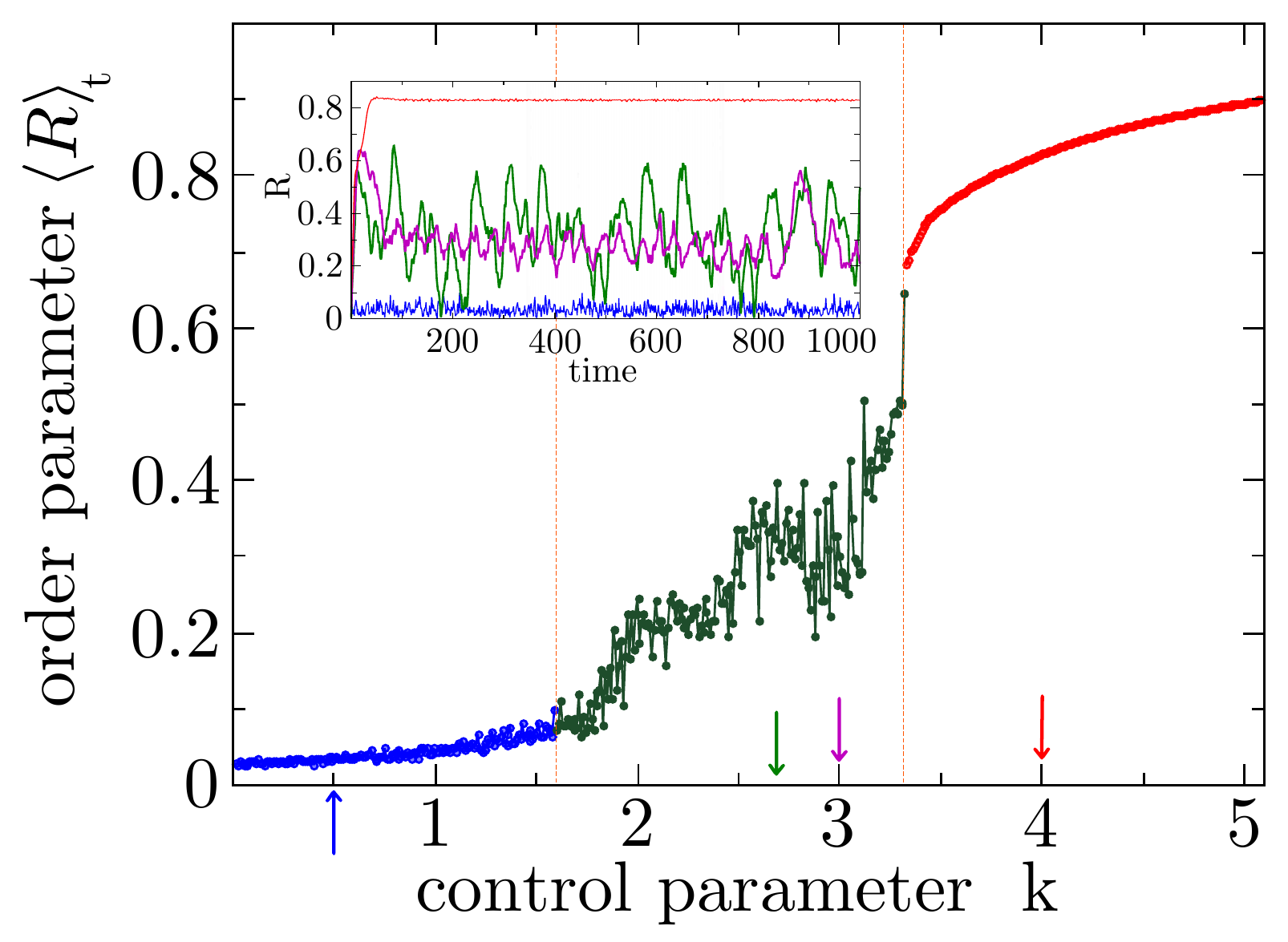}
\caption{(Main) Time-averaged value of the order parameter for the
  noisy Kuramoto dynamics running upon the Human Connectome network
  ($998$ nodes) with a Gaussian distribution of frequencies. Three
  different regimes emerge: an incoherent phase ($k < 1.6$), a
  synchronous one ($k > 3.3$), and an intermediate irregular one. In
  this last, much larger averaging times would be required to obtain
  reliable mean values and these would depend upon initial conditions,
  reflecting metastability. We have chosen not so large measuring
  times ($t=100$ for all values of $k$) to illustrate the
  variability in the intermediate region.  (Inset) Time-series for $4$
  different $k$ values, indicated by arrows in the x-axis (from left
  to right: $k=0.5, 2.7, 3.0$ and $4.0$).  }
\end{figure} 

As illustrated in Fig. 2, beside the aforementioned coherent and
incoherent phases (usually encountered in synchronizing systems) there is
an intermediate regime between them exhibiting a large variability.
Individual trajectories are depicted in the inset, for different
values of the coupling strength $k$; observe in particular the irregular
oscillations obtained for intermediate values of $k$.

The reported values of $\langle R \rangle_t $ in the main plot of Fig. 2
correspond to the time-averaged value for a single realization in its
steady state, considering up to a fixed maximum time $T$.  The
observed variability in the central region means either that (i)
larger time windows would be required for the system to self-average
or (ii) that ergodicity is broken and for each parameter value the
realization ends up in a different type of (stable or metastable)
steady state, depending on the initial condition.  This last
possibility implies that the system may remain trapped in some sort of
metastable states, from which it can escape away only after very rare
and large fluctuations.

These observations are robust against changes in the frequency
distribution, connectivity matrix normalization, and other details,
whereas the location and width of the intermediate phase are not
universal. For example, Fig. 2 has been obtained for a Gaussian
frequency distribution but similar curves are obtained for, usually
employed, Lorentzian or uniform distributions.

As this robust intermediate regime is reminiscent of Griffiths phases
in networks --posed in between order and disorder and emerging from
rare-region effects \cite{Vojta-Review,GPCN,Nat-Comm}-- it is natural
to wonder how the structural network modularity affects
synchronization dynamics in general.  As a matter of fact, it is
straightforward to convince oneself that any network consisting of
perfectly isolated moduli, each of them synchronized at different
intrinsic frequencies and phases, should exhibit oscillations of the
collective order parameter, $R$, and these oscillations are preserved
when the moduli are weakly interconnected \cite{Villegas2014}. Thus,
in large networks without delays or other additional ingredients, time
oscillations in the global coherence are the trademark of an
underlying modular structure.

To illustrate the role played by internal network modularity on global
synchronization, Fig. 3 portraits the trajectories of the parameter
$Z(t)$ in the complex plane for different values of the control
parameter $k$, measured at different hierarchical levels: two (out of
the existing $12$) different small moduli (violet and orange curves),
the two hemispheres (red and green), and the overall brain (blue).
In the incoherent phase (panel a), the real and imaginary
parts of $Z$ fluctuate around zero at all scales in the hierarchy. On
the other hand, in the coherent phase (panel d), all nodes are
synchronized, and trajectories are circles with radii close to unity
at all hierarchical levels

A much richer behavior is found in the intermediate region: panel b
(left) illustrates a situation in which one modulus
(orange) is mostly coherent, while the other (violet) is not; however,
hemispheres and global dynamics remain mostly unsynchronized (panel b
(right)). In panel c (left), we have slightly increased the control
parameter with respect to panel b, with a subsequent
increase of the coherence for all hierarchical levels.
Interestingly, as not all moduli exhibit the same state of coherence,
chaotic-like oscillations of the order parameter are observed at the global scale.

\begin{figure}[h]
\centering
\includegraphics[width=10cm,angle=0]{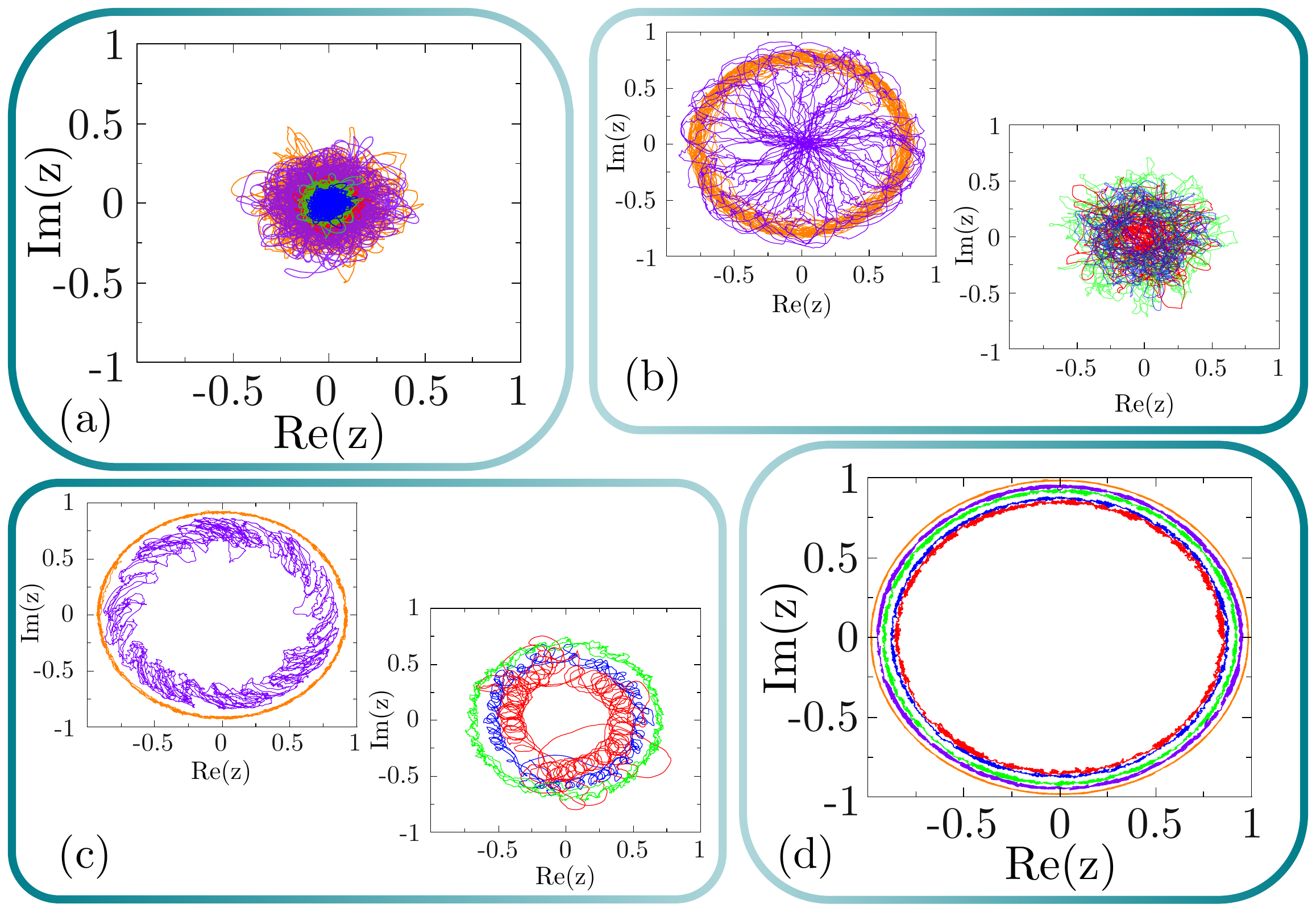}
\caption{Phase portraits of the complex order parameter $Z(t)$,
  measured at different scales in the hierarchy for a Gaussian $g(w)$
  (different realizations from those in Fig. 2): two of the
  existing moduli are plotted in violet and orange, respectively, the
  two hemispheres in red and green, and the global scale in blue.
  Panels (a)-(d) correspond to values of the control parameter
  $k=1,3,5$ and $8$, respectively (panels (b) and (c) have been split
  into two to enhance clarity). (a) In the non-ordered phase, the real
  and imaginary components of $Z$ fluctuate around zero, not
  exhibiting synchronization at any scale.  (b) In the early region of
  the intermediate phase, a few moduli are coherent (as the one in
  orange) but most of them remain unsynchronized (violet), and the
  system does not present coherence for upper scales in the
  hierarchy. (c) Increasing $k$, more heterogeneity of synchronization
  among moduli is found, and the system exhibits complex trajectories
  for the intermediate (hemispheres) and global scale. (d) In the
  coherent phase, all moduli are synchronized, and trajectories are
  concentric circles.}
\end{figure} 

We are interested in quantifying the observed variability of $R$ in
the intermediate phase. To this end, we take a particular realization
of frequencies (extracted from a Gaussian $g(w)$) and, starting from
an initial --uniformly distributed-- random configuration of
individual phases, $\{\theta_i(t=0)\}_{i=1}^N$, we measure the
temporal standard deviation of the global coherence parameter $R$
(after the transient) up to a maximum time $T=10^4$,
\begin{equation}
\sigma=\left(\langle \left(R - \langle R \rangle_t\right)^2
  \rangle_{t}\right)^{1/2}
\end{equation}
as a function of the coupling strength $k$.
\footnote{Notice that this definition of $\sigma$, that we call, ``\emph{time
variability}'' is closely related to the chimera index introduced by
Shanahan \cite{Shanahan2010}. While chimera indices are averaged
between individual network moduli and measure the onset of local
coherence, $\sigma$ is defined at the global level and records
fluctuations of the global order parameter.}  
  
As ergodicity may be broken, different initial conditions may lead to
different attractors of the dynamics, therefore we also average
$\sigma$ over $100$ different independent realizations of the
dynamical process. Results are illustrated in Fig. 4, in which we also
have plotted the diagram of the order parameter obtained for this
particular realization of $g(\omega)$ averaged over the $100$
realizations. Let us stress the following salient aspects: {\it i)}
averaged time variabilities are small in the non-coherent
($k\lesssim1$) as well as in the coherent ($k\gtrsim5$) phases,
whereas much larger variabilities are found in the intermediate region
($1\lesssim k \lesssim5$); {\it ii)} the curve of time variabilities presents
several peaks for the intermediate region, lying in the vicinity of
values of the control parameter at which the system experiences a
change in its level of coherence (see the corresponding jumps in the
derivative of the order parameter); and finally, {\it iii)} error bars
are also larger in the intermediate phase; this variability of
time variabilities means that different initial conditions can lead to
different types of time-series, suggesting a large degree of
metastability in the intermediate regime.

\begin{figure}[h]
\centering
\includegraphics[width=9cm,angle=0]{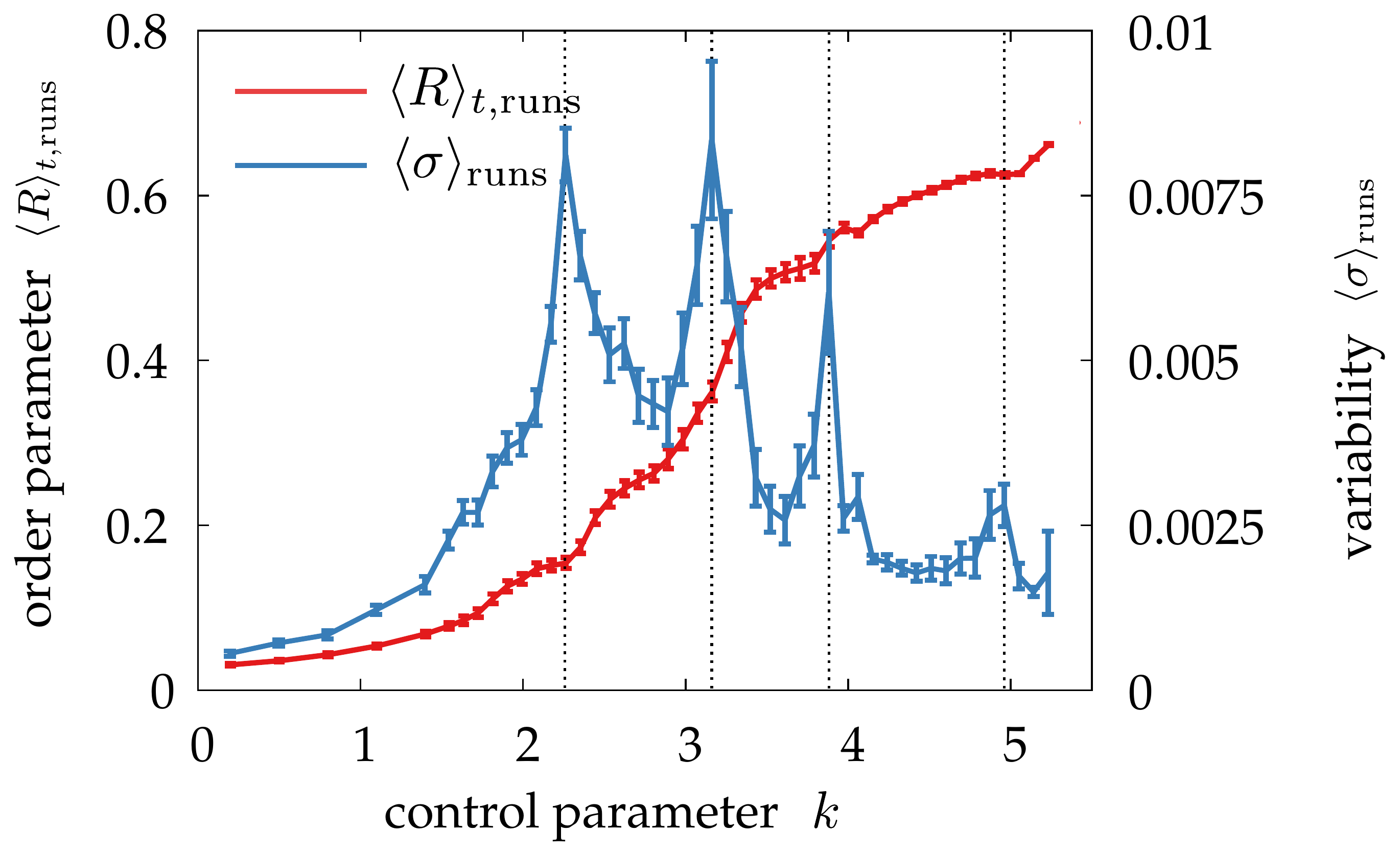}
\caption{Time-averaged order parameter $R=\langle R(t)\rangle$ and
  standard deviation of time-series, averaged over realizations
  with different --uniformly distributed-- initial conditions. Maximal
  variability is found in the intermediate phase, where the system is
  neither too unsynchronized nor too coherent. Several peaks in the
  variability can be distinguished (dashed lines), which appear at
  values of the control parameter $k$ for which the system experiments
  a fast increase in global synchronization. Statistical sampling of
  different realizations indicate that errorbars are larger in the
  intermediate region, suggesting the existence of several attractors
  depending on the initial conditions. We have averaged $100$
  different realizations, each one integrated for $10^4$ time
  steps. }
\end{figure}

\subsection{Metastability in HMNs}
Our previous results vividly illustrate the existence of an
intermediate region in which the HC exhibits maximal dynamical
variability at the global scale, suggesting metastable behavior.  In
order to explore more directly whether metastable states exist, we now
assess if the dynamics may present different attractors and, for some
values of the control parameter $k$ and noise amplitudes, if the
system may switch between different global attractors with different
levels of coherence.

Fig. 5 (a) shows a time series of the global parameter, for a fixed
realization of internal frequencies and initial phases.  It clearly
illustrates how the HC spontaneously switches between two different
attractors. These type of events, however, are not easy to observe in
the HC network. Due to the coarse-grained nature of the HC mapping,
different attractors may actually have comparable average values of
the coherence $R$, which makes their discrimination especially
difficult at the global scale.

Instead, such events are easier to spot in synthetic hierarchical
modular networks (HMN), such as proposed to model brain networks in an
efficient way (see \cite{Nat-Comm} and references therein). In such
hierarchical networks, the effects of modularity and hierarchy are
much enhanced, as they develop across a larger number of hierarchical
levels than the one allowed by current imaging techniques for
empirically obtained connectomes.

All the previously reported phenomenology is still present in such
HMNs (see \cite{Villegas2014}); in particular, the phase diagram of
the synchronization order parameter exhibits a phase transition with
an intermediate region, where variability is much enhanced
\cite{Villegas2014}. Fig. 5 (b) illustrates the bi-stable nature of
the global parameter in the intermediate phase for a HMN, in which
metastability can be very well appreciated.  This switching behavior
closely resembles ``up and down'' states, which are well known to
appear in certain phases of sleep or under anaesthesia (see
\cite{Jordi} and refs. therein).

\begin{figure}[h]
\centering\includegraphics[width=13cm]{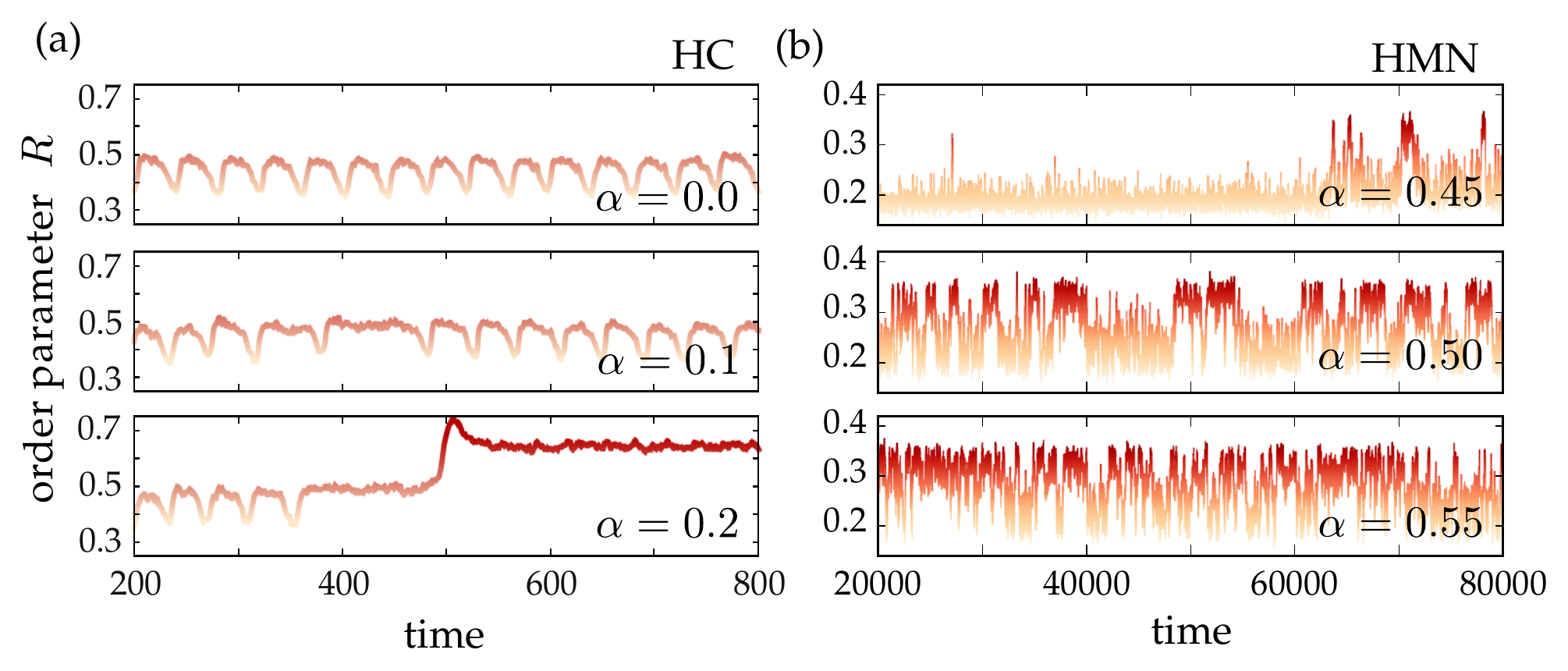}
\caption{Time series exhibit metastability of the global
  synchronization in the Human Connectome and in HMNs, in the
  intermediate region.  (a) Times series of the noisy Kuramoto
  dynamics in the HC with Gaussian $g(w)$ in the intermediate region: for low
  noise amplitudes ($\alpha=0$ and $0.1$), the system stays stable in
  the same attractor. But, for sufficiently large values, such as
  $\alpha=0.2$, the system is able to ``jump'' to another more
  coherent attractor, where it settles. (b) In HMNs (size $N=1024$,
  $9$ hierarchical levels), we observe the same phenomenology, but
  much enhanced: when noise is very low ($\alpha\leq0.45$), the system
  tends to remain stable in a certain attractor (with a few exceptions
  after very large waiting times).  Choosing a higher $\alpha$
  ($\alpha\geq0.5$), the system exhibits bi-stable behavior, switching
  intermittently between two different attractors. For large enough
  $\alpha$ ($\alpha\geq0.55$), the dynamics becomes too erratic to
  appreciate metastability. Here, frequencies were
  extracted from a Lorentz distribution.  }
\label{fig:bistability}
\end{figure}

We hypothesize that hierarchical modular networks in general (and the
HC in particular) enable the possibility of a large repertoire of
attractors, with different degrees of coherence and stability.  Such
metastability can be made evident and quantified by performing the
following type of numerical test.  Starting from a fixed random
initial condition and considering a vanishing noise amplitude
(i.e. $\alpha=0$), the system might deterministically fall into a
number of different attractors, each of them with an associated value
of the global coherence depending on the initial conditions, the
network structure, and the choice of natural frequencies. Once this
attractor A is reached, the system is perturbed by switching on a
non-vanishing noise amplitude ($\alpha >0$) during a finite time
window.  The system may remain stable in the same attractor A if the
noise is weak enough ($\alpha\ll1$). However, if larger values of the
noise amplitude are chosen, the system may jump into another close,
more stable, attractor.  If the noise amplitude is very large
($\alpha\gg 1$), the system can in principle jump to any attractor,
but, very likely, will also escape from it, wandering around a large
fraction of the configuration space. After the perturbation
time-window is over, we let the system relax once again, and check if
the new resulting steady steady state B has changed with respect to
A. In that case, we can conclude that the systems was in a metastable
state A before the perturbation, and has reached another state B after
it -- potentially a metastable state itself.

We have carried out this type of test using an artificial HMN
(see Fig. 6) for a specific value of the control parameter, $k$,
belonging in the intermediate region. Natural frequencies are sampled
from the a Lorentzian distribution $g(\omega)$ (as above, our main
results are not sensible to this choice).  Starting from a random
initial configuration of phases, we integrate Eq. \ref{eq:Kuramoto}
up to time $500$ with $\alpha=0$. After this, we introduce the
external perturbation by switching the noise coefficient $\alpha$ to a
certain non-zero value during a time window of duration $100$. Finally
we revert to $\alpha=0$ and continue the integration up to time
$t=1000$. The last steady state value is averaged over $10^{4}$
realizations of initial conditions, networks, and intrinsic
frequencies.

As illustrated in Fig. 6, for low as well as for high values of the
noise amplitude, the system has the same average order parameter close
to $\langle R \rangle_{t,\mathrm{runs}} \simeq 0.2$, as could have been
anticipated. However, a resonant peak emerges for intermediate values
of the noise, where the system switches to states with different
levels of coherence. This plot explicitly illustrates the existence of
metastability and noise-induced jumps between attractors. As noise is
enhanced, progressively more stable states are found, but above some
noise threshold, the system does not remain trapped in a single
attractor but jumps among many, resulting in a progressive decrease of
the overall coherence.

\begin{figure}[h]
\centering\includegraphics[width=9cm]{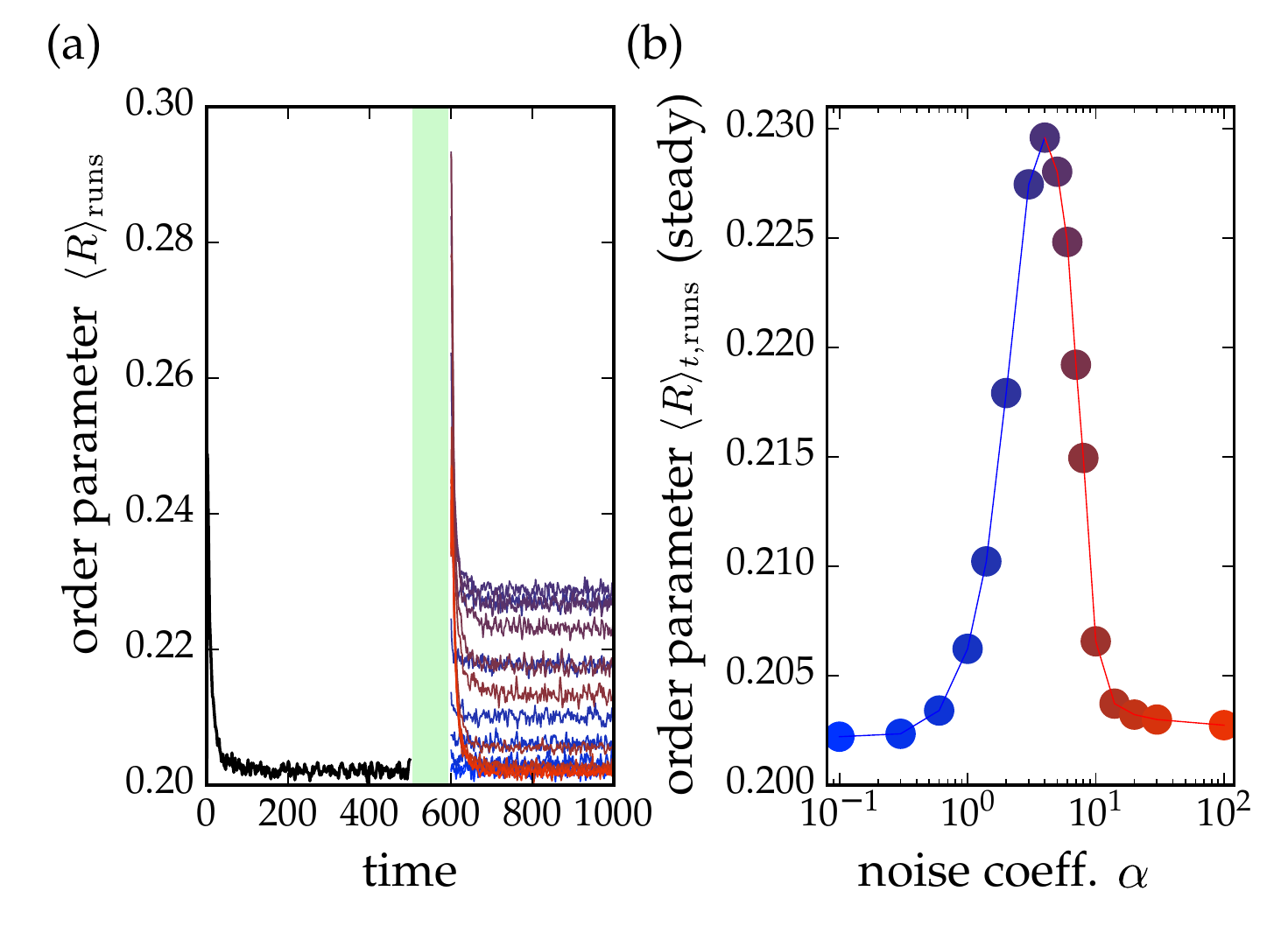}
\caption{Perturbations can lead the system to more coherent attractors
  in the intermediate non-coherent phase. (a) Order parameter $R$
  averaged in time over $10^{4}$ realizations. A noise pulse of
  amplitude $\alpha$ is applied during the green interval. This same
  protocol is repeated for different values of $\alpha$. (b) Average
  order parameter in the final steady state (after the noise pulse) as
  a function of $\alpha$. For intermediate values of $\alpha$, a
  resonant peak emerges for $1<\alpha<10$, illustrating that the
  system can jump to a close, more coherent on-average
  attractor. Simulations are run on HMN networks of size $N=1024$,
  with $9$ hierarchical levels.}
\label{fig:ordering}
\end{figure}

\section{Discussion}
It is well established that in the absence of frequency dispersion,
the Kuramoto dynamics leads to a perfectly coherent state, which is
progressively achieved in time by following a bottom-up ordering
dynamics in which increasingly larger communities become synchronized
\cite{Arenas_scales}.  

If a hierarchical modular networks is loosely connected, this type of
``matryovska-doll'' synchronization process is constrained at all
levels by structural bottlenecks, bringing about anomalously-slow
synchronization dynamics as recently reported \cite{Villegas2014}.

In the presence of intrinsic frequency dispersion the above slow
ordering process is further frustrated \cite{Villegas2014}.  For small
values of $k$ the system may remain trapped into metastable states in
which the loose connectivity between some moduli does not allow them
to overcome intrinsic-frequency differences and achieve coherence.
While persistence in metastable states may extend indefinitely,
experimental evidence suggests that the brain is able to switch
between a rich repertoire of attractors
\cite{Chialvo10,Deco2012,Haimovici}. We have shown that a simple
description of neural coherence dynamics based on the noisy Kuramoto
model may suffice to reproduce a very rich phenomenology, in
hierarchical modular networks and in particular in the human
connectome.  The introduction of small fluctuations (exemplifying
external perturbations, stimuli, or intrinsic stochasticity) allow the
system to escape from metastable states and sample the configuration
space, proving a paradigmatic modeling tool for the attractor surfing
behavior suggested by experiments. Additional ingredients, such as
explicit phase frustration \cite{Shanahan2010} or time delays
\cite{Sporns2011,Shanahan2012}, should only add complexity to the
structural frustration effect reported here, providing a finer
description of brain activity.

\section*{Acknowledgement}
We acknowledge financial support from J. de Andaluc{\'\i}a
P09-FQM-4682 and the Spanish MINECO FIS2012-37655-C02-01 and
FIS2013-43201-P.



\begin{thebibliography}{10}
\providecommand{\url}[1]{\texttt{#1}}
\providecommand{\urlprefix}{URL }

\bibitem{Hagmann}
Hagmann, P., Cammoun, L., Gigandet, X., Meuli, R., Honey, C.J., Wedeen, V.J.,
  Sporns, O.: {Mapping the structural core of human cerebral cortex.} PLoS Biol.
   6(7),  e159 (2008)
   
\bibitem{Honey09}
Honey, C.J., Sporns, O., Cammoun, L., Gigandet, X., Thiran, J.P., Meuli, R.,
  Hagmann, P.: {Predicting human resting-state functional connectivity from
  structural connectivity}. Proc. Natl. Acad. Sci.  106(6),  2035--2040 (2009)

\bibitem{Bullmore-Sporns}
Bullmore, E., Sporns, O.: {Complex brain networks: graph theoretical analysis
  of structural and functional systems}. Nat. Rev. Neurosci.  10,  186--98
  (2009)

\bibitem{Sporns}
Sporns, O.: {Networks of the Brain}. MIT Press, Cambridge (2010)

\bibitem{Review-Kaiser}
Kaiser, M.: {A tutorial in connectome analysis: topological and spatial
  features of brain networks}. NeuroImage  57(3),  892--907 (2011)

\bibitem{Review-Bullmore}
Meunier, D., Lambiotte, R., Bullmore, E.: {Modular and hierarchically modular
  organization of brain networks}. Front. Neurosci.  4,  200 (2010)  

\bibitem{Zhou06}
Zhou, C., Zemanova, L., Zamora-L{\'o}pez, G., Hilgetag, C.C., Kurths, J.: {Hierarchical
  Organization Unveiled by Functional Connectivity in Complex Brain Networks}.
  Phys. Rev. Lett.  97(23) (2006)

\bibitem{Ivkovic}
Ivkovi{\'c}, M., Amy, K., Ashish, R.: {Statistics of Weighted Brain Networks
  Reveal Hierarchical Organization and Gaussian Degree Distribution}. PLoS ONE
  7(6),  e35029 (2012)
  
\bibitem{Zhou07}
Zhou, C., Zemanova, L., Zamora-L{\'o}pez, G., Hilgetag, C.C., Kurths, J.:
  {Structure--function relationship in complex brain networks expressed by
  hierarchical synchronization}. New J. Phys.  9(6),  178--178 (2007)

\bibitem{Kaiser07}
Kaiser, M., Goerner, M., Hilgetag, C.: {Criticality of spreading dynamics in
  hierarchical cluster networks without inhibition}. New J. Phys.  9,  110 (2007)

\bibitem{Kaiser10}
{M. Kaiser}, C.H.: {Optimal Hierarchical Modular Topologies for Producing
  Limited Sustained Activation of Neural Networks}. Front. in Neuroinform.  4,
  ~8 (2010)  

\bibitem{Nat-Comm}
Moretti, P., Mu\~noz, M.A.: {Griffiths phases and the stretching of criticality
  in brain networks}. Nat. Commun.  4, 2521 (2013)  

\bibitem{Vojta-Review}
Vojta, T.: {Rare region effects at classical, quantum and nonequilibrium phase
  transitions}. J. Phys. A  39(22),  R143--R205 (2006)  

\bibitem{GPCN}
Mu\~noz, M.A., Juh\'asz, R., Castellano, C., \'Odor, G.: {Griffiths Phases on
  Complex Networks}. Phys. Rev. Lett.  105,  128701 (Sep 2010)  
  
\bibitem{Niebur2000}
Steinmetz, P.N., Roy, A., Fitzgerald, P.J., Hsiao, S.S., Johnson, K.O., Niebur,
  E.: {Attention modulates synchronized neuronal firing in primate
  somatosensory cortex}. Nature  404(6774),  187--190 (2000)
  
\bibitem{Kandel00}
Kandel, E.R., Schwartz, J.H., Jessell, T.M.: Principles of Neural Science.
  McGraw-Hill, New York (2000)

\bibitem{Buzsaki}
Buzs\'aki, G.: {Rhythms of the Brain}. Oxford University Press, NY (2006)  
      
\bibitem{Breakspear-multiscale}
Breakspear, M., Stam, C.J.: Dynamics of a neural system with a multiscale
  architecture. Phil. Trans. R. Soc. Lond. B  360(1457),  1051--1074 (2005)

\bibitem{Villegas2014}
Villegas, P., Moretti, P., Mu{\~n}oz, M.A.: Frustrated hierarchical
  synchronization and emergent complexity in the human connectome network.
  Sci. Rep.  4, 5990 (2014)

\bibitem{RPK-book}
Rosenblum, M.G., Pikovsky, A., Kurths, J.: {Synchronization -- A universal
  concept in nonlinear sciences}. Cambridge University Press, Cambridge (2001)
  
\bibitem{Kuramoto75}
Kuramoto, Y.: Self-entrainment of a population of coupled nonlinear
  oscillators. Lect. Not. in Phys.  39,  420--422 (1975)

\bibitem{Strogatz00}
Strogatz, S.H.: From kuramoto to crawford: exploring the onset of
  synchronization in populations of coupled oscillators. Physica D:  143(1),
  1--20 (2000)
  
\bibitem{Acebron-Review}
{Acebr\'on}, J.A., {Bonilla}, L.L., {P\'erez-Vicente}, C.J., {Ritort}, F.,
  {Spigler}, R.: {The Kuramoto model: A simple paradigm for synchronization
  phenomena}. Rev. Mod. Phys.  77,  137--185 (2005)

\bibitem{Arenas-Review}
Arenas, A., D\'iaz-Guilera, A., Kurths, J., {Y. Moreno}, Y., Zhou, C.:
  Synchronization in complex networks. Phys. Rep.  469(3),  93--153
  (2008)

\bibitem{Chialvo10}
Chialvo, D.R.: {Emergent complex neural dynamics}. Nature Phys.  6,  744--750
  (2010)

\bibitem{Deco2012}
Deco, G., Jirsa, V.K.: {Ongoing Cortical Activity at Rest: Criticality,
  Multistability, and Ghost Attractors}. Jour. of Neurosci.  32(10),
  3366--3375 (2012)  

\bibitem{Haimovici}
Haimovici, A., Tagliazucchi, E., Balenzuela, P., Chialvo, D.R.: {Brain
  Organization into Resting State Networks Emerges at Criticality on a Model of
  the Human Connectome}. Phys. Rev. Lett.  110,  178101 (2013)  

\bibitem{Sporns2011}
Cabral, J., Hugues, E., Sporns, O., Deco, G.: {Role of local network
  oscillations in resting-state functional connectivity.} NeuroImage  57(1),
  130--139 (2011)  

\bibitem{Radatools}
Duch, J., Arenas, A.: Community detection in complex networks using extremal
  optimization. Phys. Rev. E  72,  027104 (2005)  

\bibitem{Newman-Review}
Newman, M.: {The Structure and Function of Complex Networks}. SIAM Review
  45(2),  167--256 (2003)  

\bibitem{Shanahan2010}
Shanahan, M.: Metastable chimera states in community-structured oscillator
  networks. Chaos  20(1),  013108 (2010)

\bibitem{Jordi}
Eckmann, J.P., Feinerman, O., Gruendlinger, L., Moses, E., Soriano, J., Tlusty,
  T.: {The physics of living neural networks}. Phys. Rep.  449(1-3),
  54--76 (2007)  

\bibitem{Arenas_scales}
Arenas, A., D\'iaz-Guilera, A., P\'erez-Vicente, C.: {Synchronization reveals
  topological scales in complex networks}. Phys. Rev. Lett.  96,  114102 (2006)

\bibitem{Shanahan2012}
Wildie, M., Shanahan, M.: {Metastability and chimera states in modular delay
  and pulse-coupled oscillator networks}. Chaos  22(4),  043131+ (2012)



\end{thebibliography}

\end{document}